\documentclass[aps,twocolumn,showpacs,superscriptaddress,nofootinbib]{revtex4-1}

\usepackage{graphicx}
\usepackage{dcolumn}
\usepackage{bm}
\usepackage[bbgreekl]{mathbbol}
\usepackage{amsmath,amssymb}
\usepackage{lipsum, babel}
\usepackage{color}
\usepackage{verbatim}
\usepackage{breqn}
\usepackage{xcolor}
\usepackage[utf8]{inputenc}

\def\g2R{g^{(2)}_R}

\def\w0{w_0}

\begin{document}

\title{Nanoscale continuous quantum light sources based on driven dipole emitter arrays}

\author{Raphael Holzinger}
\affiliation{Institut f\"{u}r Theoretische Physik$,$ Universit\"{a}t Innsbruck$,$ Technikerstr. 21a$,$ A-6020 Innsbruck$,$ Austria}
\email{Raphael.Holzinger@uibk.ac.at}

\author{Maria Moreno-Cardoner}
\affiliation{Institut f\"{u}r Theoretische Physik$,$ Universit\"{a}t Innsbruck$,$ Technikerstr. 21a$,$ A-6020 Innsbruck$,$ Austria}
\affiliation{Departament de Fisica Quantica i Astrofisica and Institut de Ciencies del Cosmos$,$ Universitat de Barcelona$,$ Marti i Franques 1$,$ E-08028 Barcelona$,$ Spain.}
\author{Helmut Ritsch}
\affiliation{Institut f\"{u}r Theoretische Physik$,$ Universit\"{a}t Innsbruck$,$ Technikerstr. 21a$,$ A-6020 Innsbruck$,$ Austria}
\date{\today}

\begin{abstract}
Regular arrays of two-level emitters at distances smaller that the transition wavelength collectively scatter, absorb and emit photons. The strong inter-particle dipole coupling creates large energy shifts of the collective delocalized excitations, which generates a highly nonlinear response at the single and few photon level. This should allow to implement nanoscale non-classical light sources via weak coherent illumination. At the generic tailored examples of regular chains or polygons we show that the fields emitted perpendicular to the illumination direction exhibit a strong directional confinement with genuine quantum properties as antibunching. For short interparticle distances superradiant directional emission can enhance the radiated intensity by an order of magnitude compared to a single atom focussed to a strongly confined solid angle but still keeping the anti-bunching parameter at the level of $g^{(2)}(0) \approx 10^{-2}$.
\end{abstract}

\maketitle

\section{Introduction}

A coherently driven single two-level quantum emitter is well known to radiate non-classical light, which shows perfect anti-bunching \cite{leuchs1986photon,kimble1977photon,paul1982photon} as well as squeezing \cite{collett1984spectrum,walls1981reduced}. It is, however, very difficult to efficiently harness these properties directly. Usually it requires complex optical elements such as high-Q cavities, high aperture lenses \cite{Tey_2008,Bruno_2019} or tailored optical structures \cite{takahashi2013integrated}. Using larger ensembles of identical emitters increases the radiative output power but the nonclassical properties typically average out to create narrow band radiation but with close to thermal statistics \cite{hennrich2005transition,wolf2020light}. For small interparticle dipole interactions in a dilute gas, the dynamics of the whole ensemble in the low excitation regime then can be mapped to an effective harmonic oscillator using the Holstein Primakoff transformation \cite{ressayre1975holstein,kuzmich1997spin}.

However, recent experimental advances allow implementing and controlling precise arrays of individual quantum emitters at very close distances on the wavelength scale \cite{rui2020subradiant,bakr2009quantumgas,sherson2010single,weitenberg2011single}. Here, collective radiation effects as sub- and super-radiance play a central role and exhibit a wealth of new physical phenomena \cite{dicke1954coherence,haroche1982superradiance, guerin2016subradiance,Solano2017superradiance}. These unusual radiative properties of sub-wavelength structures of dipole coupled quantum emitters have recently been studied theoretically in great detail in various contexts \cite{temnov2005superradiance,zoubi2008bright,Porras2008collective,scully2009collective,Jenkins2012Controlled,Jenkins2013metamaterial,scully2015single,plankensteiner2015selective,bettles2015cooperative,Tudela2015Subwave,sutherland2016collective,bettles2016cooperative,bettles2016enhanced,shahmoon2016cooperativity,asenjo2016,asenjo2017exponential,ruostekoski2017arrays,hebenstreit2017subradiance,Chang2018colloquium,cottier2018role,PineiroOrioli2019dark,zhang2019theory, kornovan2019extremely,Zhang2020subradiant,Zhang2020universal,PineiroOrioli2020subradiance}. This leads to several suggestions for novel platforms for light matter coupling surpassing current limitations of quantum information protocols \cite{jenkins2016many,asenjo2017exponential,manzoni2018optimization}, precision spectroscopy \cite{ostermann2014protected,henriet2019critical} or opto-mechanics \cite{shahmoon2020cavity,shahmoon2020quantum}. Some first experimental confirmations of such collective effects were also experimentally observed, where superradiance proved more accessible than subradiance \cite{guimond2019subradiant}.

Recently we have predicted that a ringlike sub-wavelength structure with a single atom pumped at its center providing for gain can be tuned to emit spatially and temporarily coherent light \cite{holzinger2020nanoscale}. Using other operating parameters the system is also predicted to generate non-classical light with strong photon anti-bunching. In closely related foundational work it was pointed out that scattering a simple plane wave off a regular dipole array with strong dipole-dipole interaction is already sufficient to tailor the quantum statistical properties of the scattered photons \cite{williamson2020superatom,williamson2020optical}. The emerging non-clasical radiative properties here can strongly vary depending on whether the excitation frequency is tuned to superradiant or subradiant collective excitations.

\begin{figure*}[ht]
 \centering

\includegraphics[width=0.99\textwidth]{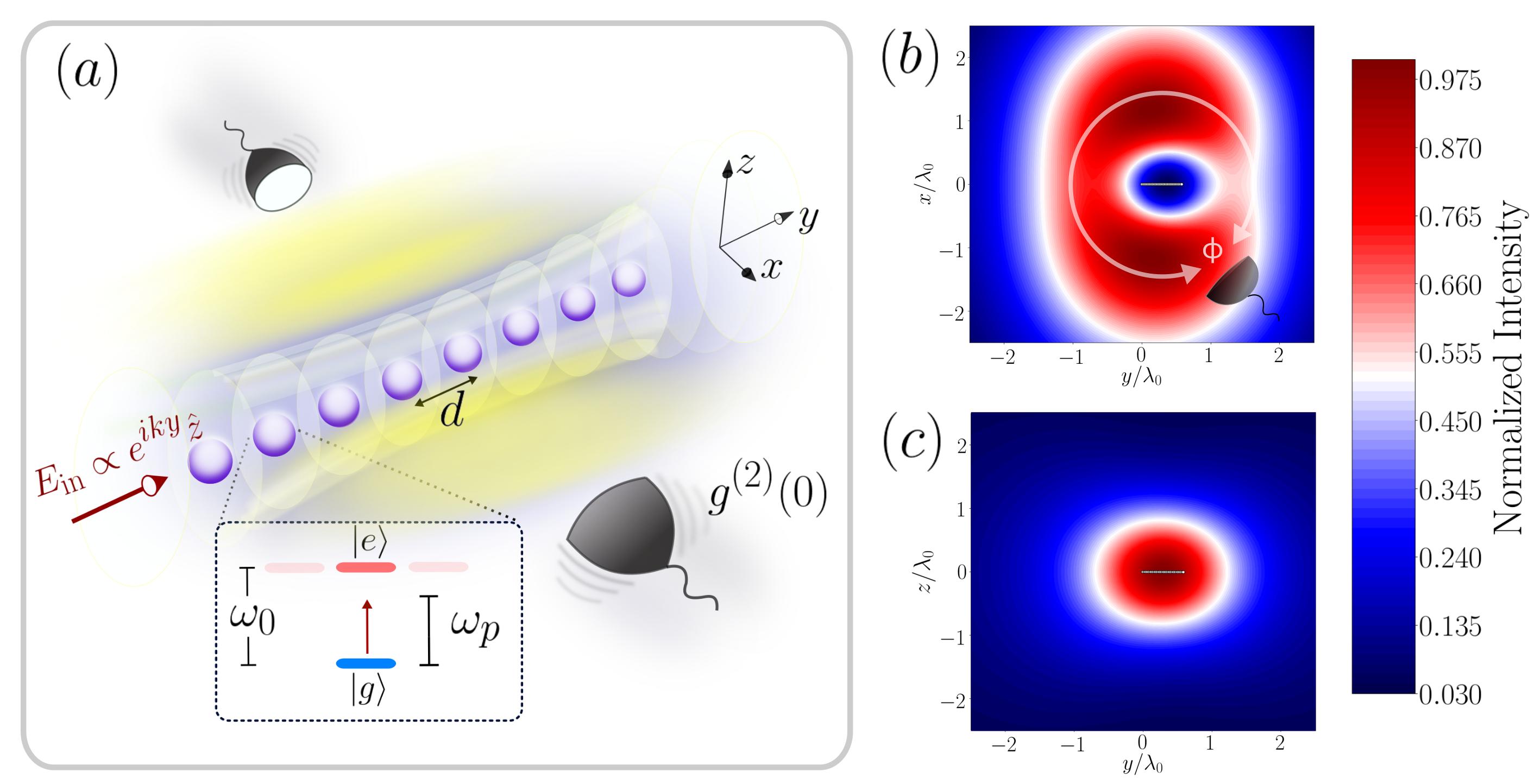}
\caption{(a) Scheme of a regular linear chain of quantum emitters with transition frequency $\omega_0$ and spontaneous emission rate $\Gamma_0$ trapped with spacing $\mathrm{d}$ along the y axis and the transition dipole moment pointing perpendicular to the chain in z direction. A coherent drive of frequency $\omega_p$ impinging from the $y$ direction with rate $\Omega_p$ along the chain is applied to all emitters and the scattered light is detected perpendicular to the chain in the far field $x$ direction. The laser is assumed to be linear polarized in z direction thus allowing to model the emitters as two-level systems only. (b) The normalized electric field intensity distribution in steady state for a chain of 30 quantum emitters with $d = \lambda_0 /40$ spacing, each being linear polarized in z direction perpendicular to the chain. The laser is tuned to the most superradiant single excitation state with a driving rate $\Omega_p = \Gamma_0$ and the cut in $z$ direction is taken at $z = 2.5\lambda_0$. Detection is done in the $xy$-plane with the detector positioned in the far field and varied around the chain with angle $\phi$. The preferred angle is in the direction of maximal scattering which simultaneously does not coincide with the laser beam direction. {(c)} The same geometry as in (b) and the cut is taken in $x$ direction at $x = 2.5 \lambda_0$.}
\label{sketch}
\end{figure*}

Here we extend such studies to explore the potential of ordered structures of dipolar quantum emitters as the basis to implement minimal non-classical light sources. In particular we study how size, geometric shape of the structure and the orientation and polarization of the individual dipoles can be used to tailor the spatial and temporal distribution of the scattered radiation as well as its genuine quantum properties. As has been seen even for a single atom \cite{vanEnk_2000,Goncalves_2020,wolf2020light}, quantum properties of the scattered radiation depend on the emission direction. In tailored larger ensembles this can be tailored to direct a large fraction of the emission towards a small angle, still keeping its special quantum statistical properties. This could e.g. point towards a high flux directional single photon source of subwavelength dimensions.

Note that the commonly used classical coupled dipole model \cite{Lee2016Stochastic} is not sufficient to account for and describe these special radiation properties, as the origin is tied to the restriction of excitations to low energy manifolds due to the strongly varying collective exciton shifts. This suppresion of multiple exciation could also be at the heart of recent obervations of antibunching in biological light harvesting structures \cite{wientjes2014strong}.

For the sake of simplicity we will mainly focus on the two cases of an ordered linear string or a regular polygon of dipoles.

This work is organized as follows. After an introduction and short review of our quantum coupled dipole model we study the spatial distribution of the emitted radiation as a function of size of the linear chain of emitters and relate it to the directional $g^{(2)}(0)$-function. In the final part we then concentrate on the non-classical aspects of the photon statistics which is closely tied to the strong nonlinear suppression of the ensemble excitations to the second and third excitation manifolds. In the supplement we discuss 2D geometries, which can confine the emission in two dimensions towards implementing higher collection efficiencies.

\section{Model}

We consider $N$ identical two-level atoms with excited state $|e\rangle$ and ground state $|g \rangle$ each, separated in frequency by $\omega_0$ with an inter-atomic distance $d \lesssim \lambda_0 = 2\pi c/ \omega_0$. The emitters are coherently pumped by a laser at rate $\Omega_p$, polarization $\pmb{\epsilon}_p$ and detuning $\Delta_p$ with respect to $\omega_0$. The corresponding raising (lowering) operators of the $i$th atom are $\sigma^{+(-)}_i$ for $i \in \lbrace 1, 2, \ldots, N \rbrace$, its dipole orientation is denoted by $\pmb{\mu}_i$ with $|\pmb{\mu}_i| = 1$ and the positions are fixed at $\pmb{r}_i$. The excited state is subject to spontaneous emission with rate $\Gamma_0$. At the considered distances, the fields emitted by each of the atoms interfere resulting in effective dipole-dipole interactions~\cite{lehmberg1970radiation}. Using standard quantum optical techniques~\cite{gardiner2004quantum} we obtain a master equation for the internal dynamics of the emitters, $ \dot{\rho} = -\frac{i}{\hbar} \left[ \mathcal{H},\rho \right] +\mathcal{L}_\Gamma \left[\rho \right]$, where the photonic part has been eliminated.

The corresponding Hamiltonian in a frame rotating at the atomic transition frequency $\omega_0$ is

\begin{dmath}
   \mathcal{H} = \sum_j \Delta_p \sigma^+_j \sigma^-_j + \sum_{i,j:i \neq j} \Omega_{ij} \sigma^+_i \sigma^-_j +  +  \Omega_p \pmb{\epsilon}_p \cdot \sum_j \hat{\pmb{\mu}}_j \left(e^{-i \pmb{k} \cdot \pmb{r}_j} \sigma^+_j + e^{i \pmb{k} \cdot \pmb{r}_j} \sigma^-_j \right),
   \label{eq:hamiltonian}
\end{dmath}

where we have assumed that the incident field is a plane wave with wave-vector $\pmb{k}$ ($|\pmb{k}| \approx 2\pi/\lambda_0$), while the Lindblad operator accounting for collective spontaneous emission reads
\begin{equation} \label{eq:lindblad_gamma} 
\mathcal{L}_\Gamma \left[ \rho\right] = \sum_{i,j} \frac{\Gamma_{ij}}{2}\left(2 \sigma^-_i\rho \sigma^+_j -\sigma^+_i \sigma^-_j \rho - \rho \sigma^+_i \sigma^-_j \right).
\end{equation}

The collective coupling rates $\Omega_{ij}$ and $\Gamma_{ij}$ are given as the real and imaginary part of the overlap of the transition dipole moment of the $i$th atom with the electromagnetic Green's tensor by the $j$th atom.
The Green's tensor in free space acting on an oscillating unite dipole is given by

 \begin{equation} \label{green-app}
\begin{split}
        \pmb{{G}}(\pmb{r},\omega_0) \cdot \hat{\pmb{\mu}} &= \frac{e^{i k_0 r}}{4 \pi r} \Big[(\hat{\pmb{r}} \times  \hat{\pmb{\mu}}) \times  \hat{\pmb{r}} + \\
        &+\Big(\frac{1}{k_0^2 r^2}-\frac{i}{k_0 r}\Big)(3 \hat{\pmb{r}}( \hat{\pmb{r}}\cdot  \hat{\pmb{\mu}})- \hat{\pmb{\mu}})\Big].
\end{split}
\end{equation}

Here, $ \hat{\pmb{r}} = \pmb{r}/|\pmb{r}|$ is the unit vector in the direction where the Green's tensor is evaluated, $k_0 = \omega_0/c$ is the wavenumber of the emitted light by the dipole and $\hat{\pmb{\mu}} = \pmb{\mu}/|\pmb{\mu}|$ is the unit dipole orientation.

After solving for the atomic density matrix the electric field generated at position $\pmb{r}$ can be found from a generalized input-output relation which in the absence of external fields is given by\cite{hood_2017,asenjo2017exponential}
 \begin{equation} \label{e-field2}
 \hat{\pmb{E}}^+(\pmb{r})  = \mu_0 \omega_0^2 \sum_{j=1}^N \pmb{G}(\pmb{r}-\pmb{r}_j,\omega_0) \cdot \hat{\pmb{\mu}}_j \sigma^-_j.
\end{equation}

The field intensity can then be evaluated from the emitted electric field operator $\pmb{E}^+(\pmb{r})$ at position $\pmb{r}$:
\begin{equation} \label{eq:intensity} 
I(\pmb{r}) = \left \langle \pmb{E}^+(\pmb{r}) \pmb{E}^-(\pmb{r}) \right \rangle.
\end{equation}
For typical operating conditions, we show the emitted light intensity of the atomic array steady state in Fig.\ref{sketch} (b) and (c).

Similarly, we can obtain from Eq. \ref{e-field2}, normalized (zero-time delay) second-order correlation functions, defined as:

\begin{equation} \label{eq:g2}
g^{(2)}(0) = \frac {\langle \hat{\pmb{E}}^+(\pmb{r}) \hat{\pmb{E}}^+(\pmb{r})\hat{\pmb{E}}^-(\pmb{r})\hat{\pmb{E}}^-(\pmb{r})\rangle}{|\langle \hat{\pmb{E}}^+(\pmb{r})\hat{\pmb{E}}^-(\pmb{r})\rangle|^2}.
\end{equation}
We will evaluate $g^{(2)}(0)$ in the far field ($|\pmb{r}| = r \gg \lambda_0$) and as a function of the emission azimuthal angle $\phi$, defined in the plane containing the array and in the direction of maximal emission (see Fig.\ref{sketch}). The angular position as a function of $\phi$ in Figs. \ref{direction} and \ref{scan} is given by $\hat{\pmb{r}}(\phi) = (sin(\phi),-cos(\phi),0)^T$.

The steady state excited state population $\langle \hat{n}_{\mathrm{ex}}\rangle = \sum_{j=1}^N \langle \sigma^+_j \sigma^-_j \rangle$ is closely related to the photon emission rate $\Gamma_{\mathrm{out}} = \sum_{ij}^N \Gamma_{ij} \langle \sigma^+_i \sigma^-_j \rangle $ which for $N$ initially fully inverted two-level emitters is simply given by $N \Gamma_{ii} = N\Gamma_0$. For a steady state which consists only of a superposition of a single photon state $m$ and  the ground state, the total emission rate is given by $\Gamma_m \langle n_{\mathrm{ex}} \rangle$, where $\Gamma_m$ is the collective decay rate of the state $m$ and $\langle n_{\mathrm{ex}} \rangle$ the emitter population in the chain. On the other hand a single dipole in free space coherently driven by a laser at rate $\Omega_p$ and detuning $\Delta_p$ with respect to the atomic transition frequency has an effective decay rate of $\Gamma_0 \Omega_p^2/(4 \Delta_p^2 +\Gamma_0^2+2\Omega_p^2)$ which can be derived by solving the atomic master equation in the steady state. The effective rate saturates at $\Gamma_0/2$ which constitutes the maximum steady state emission rate of a single coherently driven two level emitter. The total photon emission rate of closely spaced quantum emitters as e.g. in Fig. \ref{sketch}(a) can be many times that value as is demonstrated in the following.

\section{Spatial distribution of resonantly scattered light }

Similar to~\cite{Clemens_2004,clemens_2003,Wagner_2010,garcia_2020} the directional intensity is calculated by integrating over an angular width $\Delta \phi= 0.01\pi $ covered by a detector in the far field,

\begin{equation} \label{directional}
\mathcal{J}(\phi) = \frac{\Delta \Omega}{\Delta \phi} \int^{\phi + \Delta \phi}_{\phi - \Delta \phi} \langle \hat{\pmb{E}}^+(\hat{\pmb{r}})\hat{\pmb{E}}^-(\hat{\pmb{r}})\rangle d \phi.
\end{equation}

Here $\Delta \Omega$ is the solid angle of a square detector varying over the emission angle $\Delta \phi$. The expression for the directional emission $\mathcal{J}(\phi)$ is equivalent to Eq. 5 in the case of an infinitesimal small detector.

\begin{figure}[t]
\includegraphics[width=0.48\textwidth]{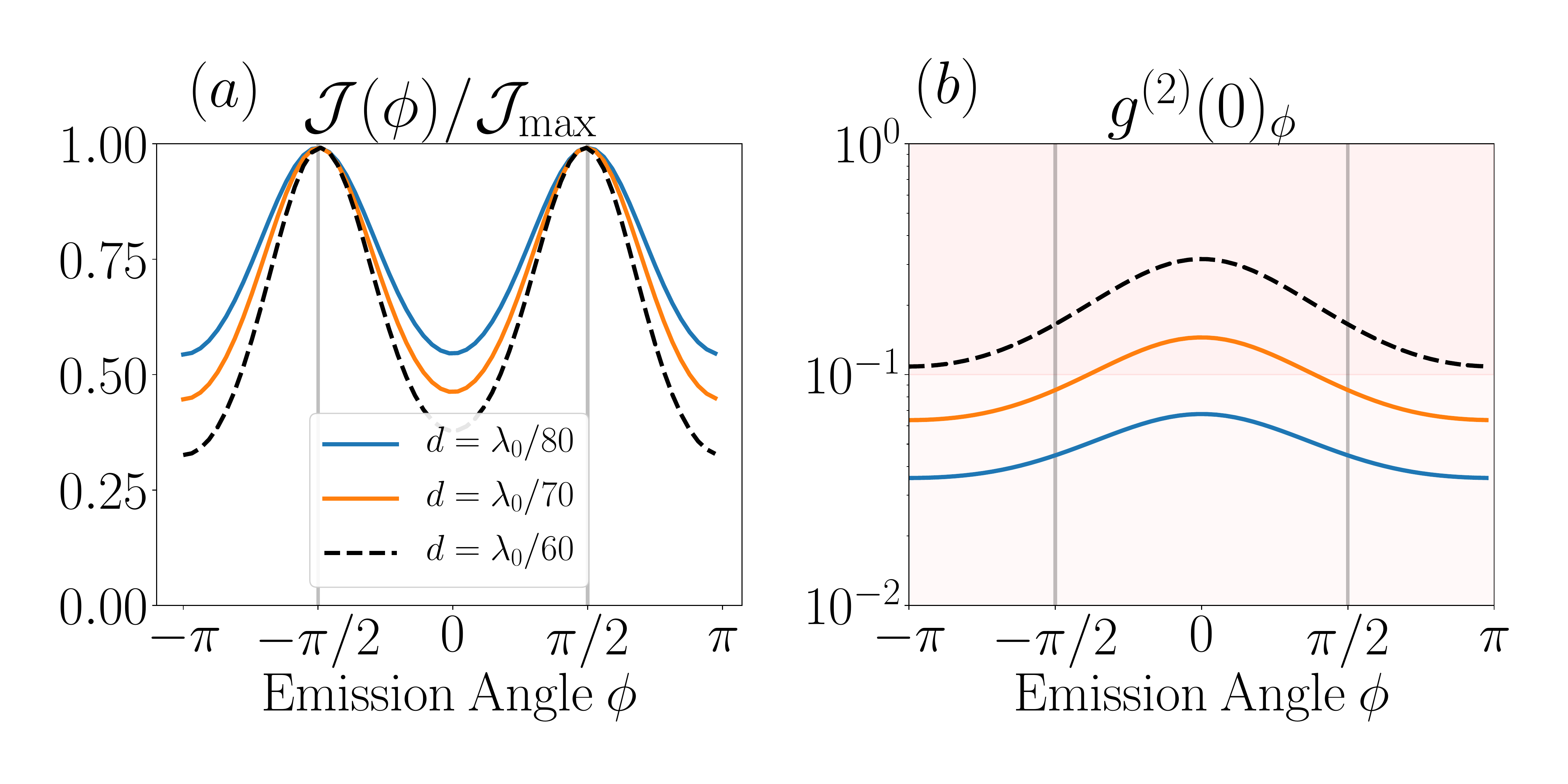}
 \vspace*{-6mm}
\caption{\textit{Steady State Emission}. Normalized angular far field intensity distribution and bunching parameter $g^{(2)}(0)|_\phi$ for a linear chain of $30$ emitters linear polarized in $z$ direction  with various lattice constants $d$ and the same laser setup as in Fig. 1(a,b). The laser beam propagates at the angle $\phi = 0$ along the chain, from $(\hat{x},\hat{y},\hat{z})=(0,-1,0)$.
{(a)} shows that maximal emission occurs perpendicular to the chain for all lattice constants which is favorable for photon detection as totally reflected light would interfere with the laser beam.
In {(b)} the corresponding bunching parameter shows that a photon bunching parameter of $g^{(2)}(0)|_\phi \le 0.1$ is achieved for a small enough lattice spacing in the direction of maximal scattering, which is perpendicular to the laser beam direction. The vertical lines at $\pm \pi/2$ indicate the angles of maximal scattering.}
\label{direction}
\end{figure}

\begin{figure}[t]
\includegraphics[width=0.48\textwidth]{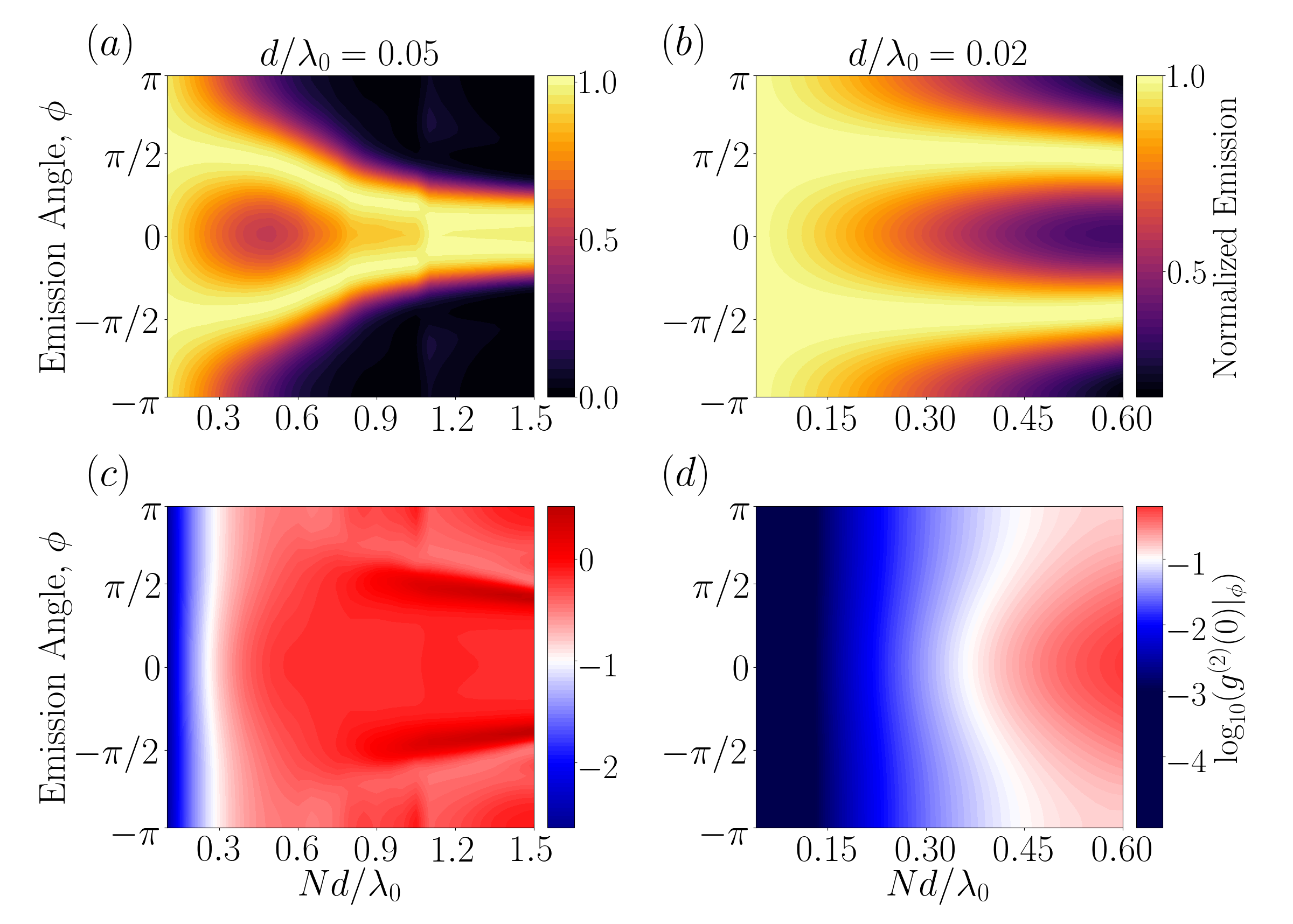}
 \vspace*{-5mm}
\caption{Normalized directional emission and the photon bunching parameter $g^{(2)}(0)$ as a function of the emission angle $\phi$ and total chain length $Nd/\lambda_0$ in the steady state. The laser parameters are identical to Fig. \ref{sketch}(b,c) with the laser beam angle at $\phi = 0$ and the lattice constants $d$ in the first and second column are $0.05\lambda_0$ and $0.02 \lambda_0$ respectively. The normalized directional emission in (a) shows initially perpendicular emission to the chain and reflection ($\phi = 0$) for a chain length $Nd\gtrsim\lambda_0$. (b) and (d) show directional emission perpendicular to the laser beam direction with a photon bunching of $g^{(2)}(0) \le 0.1$ for a lattice spacing $d=0.02 \lambda_0$. Note that the emission intensity is normalized over the emission angle for each chain length individually.}
\label{scan}
\end{figure}

\begin{figure*}[t]
\includegraphics[width=1\textwidth]{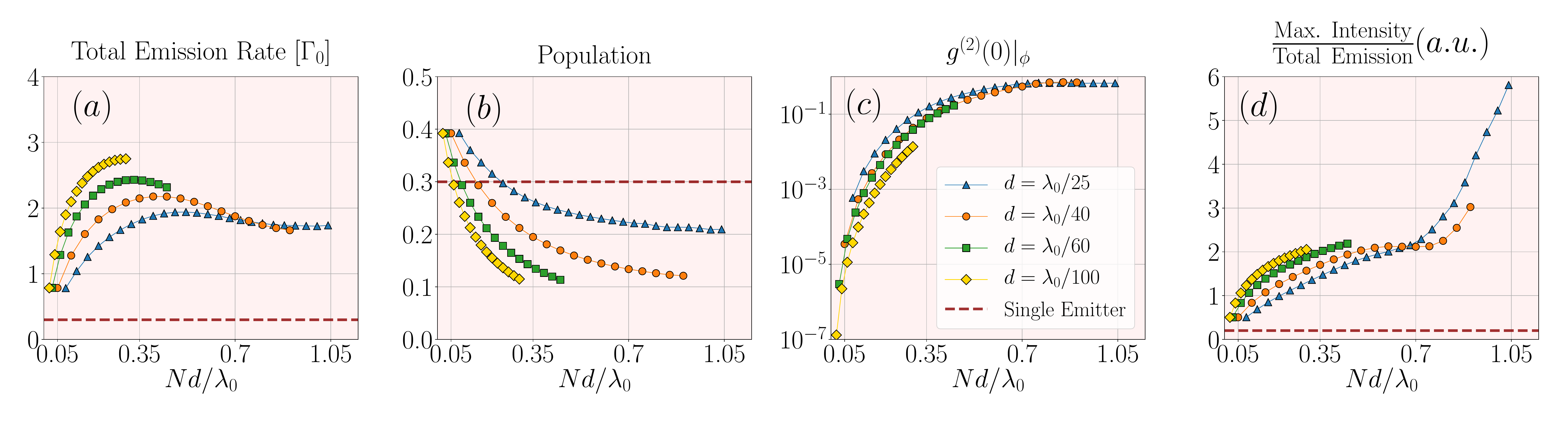}
 \vspace*{-10mm}
\caption{The steady state of a coherently driven linear chain of emitters with different lattice constants $d$ for the same geometry and parameters as in Fig 1.
In (a) and (b) the total photon emission rates and the emitter populations are shown respectively and show an increasing superradiant emission rate for decreasing $d$.
    (c) Photon correlations $g^{(2)}(0)$ are measured in the direction of maximal emission in the far field and which is as seen in Figs. 1-3 perpendicular to the chain and laser beam in $x$ direction for a total chain length of $Nd \gtrsim 0.35\lambda_0$. A bunching parameter of $g^{(2)}(0) \le 0.1$ in combination with directional perpendicular emission w.r.t. to the lasing direction is reached for sufficiently small lattice constants $d$. (d) The maximal intensity in the $xy$ plane as can be seen in Figs. 1-3 divided by the total emission rate. The sudden increase of the peak intensity for a total chain length $\gtrsim 0.7 \lambda_0$ originates from the merger of the two emission peaks as seen in Fig. \ref{scan} (a) which turns the perpendicular emission into a reflection in the laser beam direction. In comparison the dashed line shows the steady state values for a single emitter coherently driven with a rate $\Omega_p = \Gamma_0$ with the emission being uniform in the $xy$-plane without directionality. Note that in the case of maximal emission in two directions (see Fig. 1 (a) and (b)) the maximal intensity is calculated for one of the maxima.}
\label{scalings}
\end{figure*}

\begin{figure}[ht]
\includegraphics[width=0.49\textwidth]{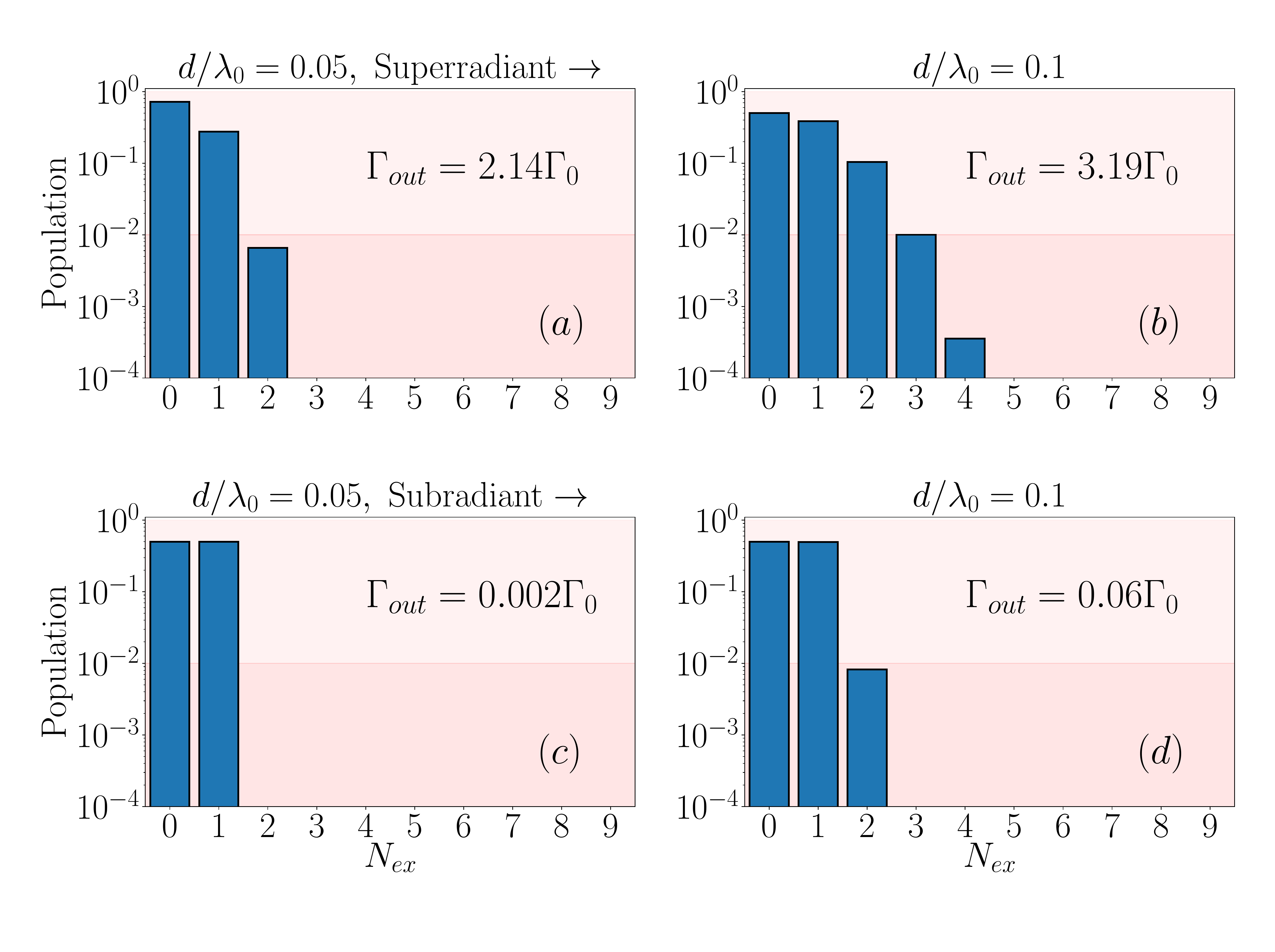}
 \vspace*{-12mm}
\caption{Excitation energy distribution in steady state for a coherently driven linear chain of $9$ emitters along the $y$ direction each linear polarized in $z$ direction with the linear polarized drive along the chain direction and pump amplitude $\Omega_p = \Gamma_0$. The $x$ axis denotes the excitation manifolds which are proportional to the output electric field (see Eq. \ref{gamma}). For {(a)} and (b) in the first row the laser is tuned to resonance with the maximally superradiant single excitation eigenstate while for {(c)} and {(d)} in the second row we target the most subradiant single excitation state in a steady state operation (In the supplement a pulsed preparation of the most subradiant state is shown). In comparison the dashed line shows the steady state values for a single emitter coherently driven with a rate $\Omega_p = \Gamma_0$ with the emission being uniform in the $xy$-plane without directionality. Note very strong (nonlinear) suppression of higher excitation numbers for short inter-particle distances. The red horiztonal shades are guides for the eyes only.}
\label{statistics}
\end{figure}

In Fig. \ref{direction}(a) the directional emission pattern for a linear chain is normalized by the maximal emission $\mathcal{J}_{\mathrm{max}}$ over the emission angle $\phi$ for various lattice constants $d$.
 The geometry is identical to Fig. \ref{sketch} with the emitters linear polarized in z direction and a linear polarized laser beam propagating along the chain direction with a pumping rate $\Omega_p = \Gamma_0$ tuned to the most superradiant single excitation eigenmode of the chain. In Fig. \ref{direction}(b) the bunching parameter $g^{(2)}(0)|_\phi$ has values $\le 0.1$ in the direction of maximal emission for $d \le \lambda_0/70$ with the emission exhibiting two emission maxima perpendicular to the laser beam direction ($\phi=0$).
In order to show the dependence of the directional emission on the total chain length we plot both the directional emission and the bunching parameter for fixed $d$ and an increasing number of emitters in Fig. \ref{scan} for the same parameters as in Fig. \ref{direction}.
The emission as seen in Fig. \ref{scan} (a) and (b) becomes highly directional for increasing chain lengths \cite{Zoubi_2010}. The normalized $g^{(2)}(0)|_\phi$ in (c) and (d) on the other hand shows strong anti-bunching at smaller spacings between the emitters as opposed to an overall small chain length. Directional photon statistics of this kind have been recently observed for two emitters by varying the detector position~\cite{wolf_2020}.
As described in~\cite{asenjo2017exponential, masson_2020} the $N$ single excitation eigenstates of a linear chain are spin waves with a well defined wave-vector and corresponding eigenfrequencies. For finite chains the spin wave solution is only approximately valid but still provides a qualitatively insightful picture in order to better understand the energy spectrum and is discussed in the supplement.
In all examples the laser is on resonance with the most superradiant mode of these spin waves in order to maximize the photon emission rate.
To separate the scattered quantum light from the driving laser light, the laser illuminates the ensemble from the $y$ direction and is subsequently scattered at maximal intensity orthogonal to the chain in $x$ direction. Another scheme is shown in the supplementary material where a ring of linear polarized emitters in a ring geometry scatter light directionally coming from a circular polarized laser. A decrease of the emission angle with increasing emitter number has been discussed in \cite{asenjo2017exponential} for subradiant states in a linear chain and it's pulsed is discussed in the supplement as well. Similarly we show in Fig. \ref{scalings} how increasing the total length $Nd$ for various lattice spacings $d$, the light emission becomes more focussed by targeting the superradiant mode of the chain \cite{ballantine_2020,ballantine2_2020,williamson_2020,Parmee2020,Williamson2_2020}. To increase the total photon emission rate in Fig. \ref{scalings} (a) it is reasonable to increase the number of emitters, but as seen in Fig. \ref{scalings} (c), the increased chain length leads to a higher steady state $g^{(2)}(0)$ in the direction of maximal emission. But by decreasing the lattice spacing for a given chain length the $g^{(2)}(0)$ in the direction of maximal emission is decreasing as well.
As a figure of merit for directional emission the ratio between the maximal intensity in the $xy$-plane and the total emission rate is plotted Fig. \ref{scalings} (d) and as the total emission rate in (a) for $d=\lambda_0/100$ is the highest, its ratio in (d) is the highest as well, indicating that a larger amount of the totally emitted light is concentrated in the emission peak(s). The sudden increase in the ratio in (d) for a chain length of $\gtrsim 0.7 \lambda_0$ originates from the merger of two emission peaks into one, which is reflected back into the laser beam direction as seen in Fig. \ref{scan} (a).
The decrease of the emitter population plotted in Fig. \ref{scalings} (b) with increasing emission rate shows the increasing superradiance of the targeted eigenmode of the chain which possesses a decay rate of $\approx N \Gamma_0 \langle n_{\mathrm{ex}}\rangle$ for $d \le \lambda_0/100$. On the other hand the decrease of the total emission rate for increasing chain length is due to less emitter population in the superradiant mode. Fig. \ref{scalings} shows that both a directional emission and a bunching parameter $g^{(2)}(0) \le 0.1$ in the direction of maximal emission can be achieved for lattice spacings of $d \approx \lambda_0/100$ as the chain length exceeds $\lambda_0/3$. The dashed lines in Fig. \ref{scalings} show the case of a single emitter coherently driven with a rate $\Omega_{\mathrm{p}} = \Gamma_0$. The $g^{(2)}(0)$ function is zero in this case but the small emission rate and absence of emission peaks in the $xy$-plane renders the efficient collection of the scattered light less favorable. Although the distances between the emitter are out of reach for optical tweezers and lattices at present, recently \cite{hilbin2021room} superradiant and subradiant states in single nanocuboids at room temperature have been observed. Here the lattice constants of the nanocube can be adjusted and the spacing between the interacting dipoles can be even below $\lambda_0/100$.

\section{Photon statistics}

In the following, we will discuss photon statistics in terms of the occupations in the various excitation manifolds since we  have eliminated the photonic modes during the derivation of the atomic master equation, however, the output intensity is proportional to the atomic raising/lowering operators weighted by the Green's tensor. In Fig. \ref{statistics} it can be seen that single photon states of a high quality can be obtained at $d\le \lambda_0/10$ for a laser tuned both to the superradiant (in Fig. \ref{statistics} (a,b)) and subradiant (in Fig. \ref{statistics} (c,d)) eigenmode of the chain. The geometry is identical to the previous examples with the laser driving the chain from the y direction at a rate $\Omega_p = \Gamma_0$. For the subradiant case in Fig. \ref{statistics} (c) the total emission rate for a linear chain of $9$ emitters at a spacing of $0.05\lambda_0$ shows that a steady state occupation of $50 \%$ in the single excitation manifold is reached. The total emission rate indicates a lifetime $500$ times longer than that of a single excited emitter. The statistics in Fig. \ref{statistics} are obtained by continuous laser driving in the steady state and if the laser is tuned to the most subradiant eigenmode of the chain the emitter population levels out at $50 \%$ as for a chain emission rate of $\ll \Gamma_0$ a pumping rate of $\Omega_p = \Gamma_0$ resembles the case of a single strongly driven emitter. In this regime the chain prepared in the subradiant collective mode behaves like a single strongly driven emitter which reaches a excited state occupation of $0.5$ in the steady state. In the supplement we demonstrate an efficient way to prepare a subradiant eigenmode via a laser pulse.
Pulsed on-demand production of photon number states is proposed in \cite{groiseau2020deterministic} with a single multi-level atom in a cavity.

\section{Conclusions}
We have shown that the collective excitations in regular sub-wavelength structures of dipolar quantum emitters can behave like a designer two-state atom \cite{pruchyathamkorn2020complex} with a strongly enhanced effective dipole moment and a tailorable spatial radiation pattern. When illuminated with weak coherent light their strong nonlinear response at the single or few photon level leads to directional emission of strongly anti-bunched light without the help of any additional optical elements. While we restricted ourselves here to the most simple generic cases of regular chains, more general structures can be envisaged for specific applications and in particular as the basis of minimalist nanoscale single photon sources. We highlight these surprising features for the generic examples of a regular polygon with titled polarization and interacting regular polygons which are partially pumped in the supplement.

Even at fairly high excitation powers, in strongly interacting configurations the second excitation manifold is only very weakly excited, which might be a hint on the origin of anti-bunching found in bio-molecular dipole configurations \cite{wientjes2014strong}. Interestingly, as shown in the supplement for a ring with slightly tilted dipoles, besides the spatial dipole arrangement, polarization can be an efficient tool to separate excitation light and emitted photons in addition to directional spatial filtering. Note that while atoms in tweezers might be the most straightforward test bed for these ideas, alternatives used synthetic molecules \cite{pruchyathamkorn2020complex}, nanocuboids \cite{hilbin2021room} at room temperatures or quantum dot nano-structures \cite{dalacu-2019} and should largely exhibit similar physics as long as the couplings are comparable with environmental decoherence effects.

\begin{acknowledgments}

We thank David Plankensteiner for helpful discussions and acknowledge funding from the Austrian Science Fund (FWF) doctoral college DK-ALM W1259-N27 (R. H.), and the European Unions Horizon 2020 research and innovation program under the Marie Sklodowska-Curie grant agreement No. 801110 and the Austrian Federal Ministry of Education, Science and Research (BMBWF) (M.M.-C.). It reflects only the authors view and the Agency is not responsible for any use that may be made of the information it contains. Numerical simulations were performed with the Julia programming language including the QoJulia.org quantum optics package \cite{kramer2018quantumoptics}.

\end{acknowledgments}




%

\appendix

\begin{figure*}[ht]
\includegraphics[width=0.99\textwidth]{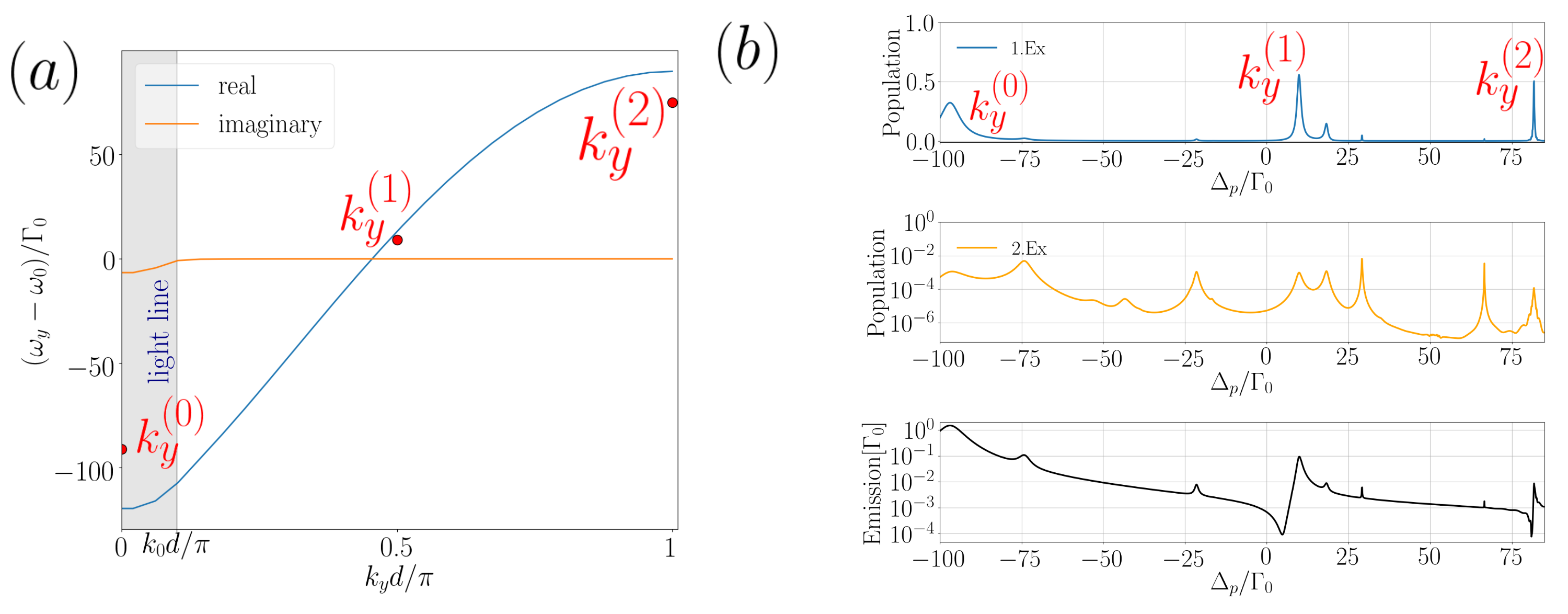}
\caption{\textit{Energy Band.} In {(a)} the continuous lines represent a chain of $N=50$ emitters whereas the dots indicate the real part of $(\omega_y-\omega_0)/\Gamma_0$ of a $5$ emitter chain. Since the dispersion relation is symmetric around the $y$-axis the first Brillouin-Zone is shown only for wavenumbers $k_y \in [0,\pi/d]$.
The spacing is $d/\lambda_0 = 0.05$ and all emitters are linearly polarized along the chain.
{(b)} shows the steady state populations and total emission rate of a $5$ emitter chain driven coherently from the $z$-direction. The $x$ axis shows the laser detuning with respect to the single atom transition frequency $\omega_0$. Resonances at the single-excitation eigenenergies are clearly visible and show significant populations in the steady state and note that the linewidths of the resonances is given by the decay rates of the targeted modes.}
\label{energy-band}
\end{figure*}

\section{Green's Tensor and effective Model}

The electric field generated by an ensemble of atoms is mediated by the electromagnetic Green's tensor, moreover in the weak excitation limit the Hamiltonian can be rewritten in a form which allows for analytic solutions of its eigenvalue equation. These steps are described below for the special case of an infinite chain. For a finite chain the procedure is only valid approximately because of its open boundaries at the ends of the chain but for an infinite chain it becomes accurate and as investigated previously\cite{masson_2020,asenjo2017exponential} 1D atomic arrays behave like a quantum waveguide and support guided modes that do not decay into free space.

The Green's tensor in free space acting on an oscillating unite dipole is given by

 \begin{equation} \label{green}
\begin{split}
        \pmb{{G}}(\pmb{r},\omega_0) \cdot \hat{\pmb{\mu}} &= \frac{e^{i k_0 r}}{4 \pi r} \Big[(\hat{\pmb{r}} \times  \hat{\pmb{\mu}}) \times  \hat{\pmb{r}} + \\
        &+\Big(\frac{1}{k_0^2 r^2}-\frac{i}{k_0 r}\Big)(3 \hat{\pmb{r}}( \hat{\pmb{r}}\cdot  \hat{\pmb{\mu}})- \hat{\pmb{\mu}})\Big].
\end{split}
\end{equation}

Here, $ \hat{\pmb{r}} = \pmb{r}/|\pmb{r}|$ is the unit vector in the direction where the Green's tensor is evaluated, $k_0 = \omega_0/c$ is the wavenumber of the emitted light by the dipole and $\hat{\pmb{\mu}} = \pmb{\mu}/|\pmb{\mu}|$ is the unit dipole orientation. The coherent and dissipative interaction rates between emitter $i$ and $j$ read

 \begin{equation} \label{omega}
 \Omega_{ij} = -\frac{3\pi \Gamma_0}{k_0} \mathrm{Re} \{ \hat{\pmb{\mu}}_i^* \cdot \pmb{G}(\pmb{r}_i-\pmb{r}_j,\omega_0)\cdot \hat{\pmb{\mu}}_j\},
\end{equation}

 \begin{equation} \label{gamma}
 \Gamma_{ij} = -\frac{3\pi \Gamma_0}{k_0} \mathrm{Im} \{ \hat{\pmb{\mu}}_i^* \cdot \pmb{G}(\pmb{r}_i-\pmb{r}_j,\omega_0)\cdot \hat{\pmb{\mu}}_j \}
\end{equation}

where $\hat{\pmb{\mu}}_i$ is the unit dipole moment associated with the transition of atom $i$.
The single atom spontaneous emission rate is given by $\Gamma_0 = \omega_0^3 |{\pmb{\mu}}|^2/3\pi \epsilon_0 \hbar c^3$.

As described in the main text after solving for the atomic density matrix the electric field generated at position $\pmb{r}$ can be obtained from a generalized input-output relation \cite{hood_2017,asenjo2017exponential}, which in the absence of external fields is given by
 \begin{equation} \label{e-field}
 \hat{\pmb{E}}^+  = \mu_0 \omega_0^2 \sum_{j=1}^N \pmb{G}(\pmb{r}-\pmb{r}_j,\omega_0) \cdot \hat{\pmb{\mu}}_j \sigma^-_j.
\end{equation}

To obtain this equation the Markovian approximation has been made, therefore retardation effects arising from the physical sepration between the quantum emitters can be ignored. As long as the ensemble stays within the length scales of $\approx 1$ meter this assumption is well founded.\cite{Chang_2012,shi_2015,thompson_1992}

We will consider the weak excitation limit in which only the single-excitation manifold is significantly occupied and neglecting external driving fields for the moment.
In this case the recycling term which is the first term in the Lindblad expression can be neglected.
This term accounts for the change in the ground state
population. Then,
the system can be fully understood from the properties
of the eigenstates of an non-Hermitian Hamiltonian that reads (setting $\hbar = 1$) $\mathcal{H} = \omega_0 \sum_{i=1}^N \sigma^+_i \sigma^-_i+\mathcal{H}_{\mathrm{eff}}$ with

 \begin{equation} \label{eq:ham-eff}
 \begin{split}
\mathcal{H}_{\mathrm{eff}} &= -\mu_0 \omega_0^2 \sum_{i,j=1}^N \hat{\pmb{\mu}}^*_i \cdot \pmb{G}(\pmb{r}_i-\pmb{r}_j,\omega_0) \cdot \hat{\pmb{\mu}}_j \sigma^+_i \sigma^-_j \\
&= \sum_{i,j=1}^N \Big(\Omega_{ij}-i \frac{\Gamma_{ij}}{2}\Big) \sigma^+_i \sigma^-_j,
\end{split}
\end{equation}

where $\Omega_{ii} = 0$, as it only leads to a finite global energy shift in the Hamiltonian for identical emitters. This model will now be used to to investigate the properties of a linear chain of atoms in free space without external driving.

\begin{figure*}[ht]
\includegraphics[width=0.9\textwidth]{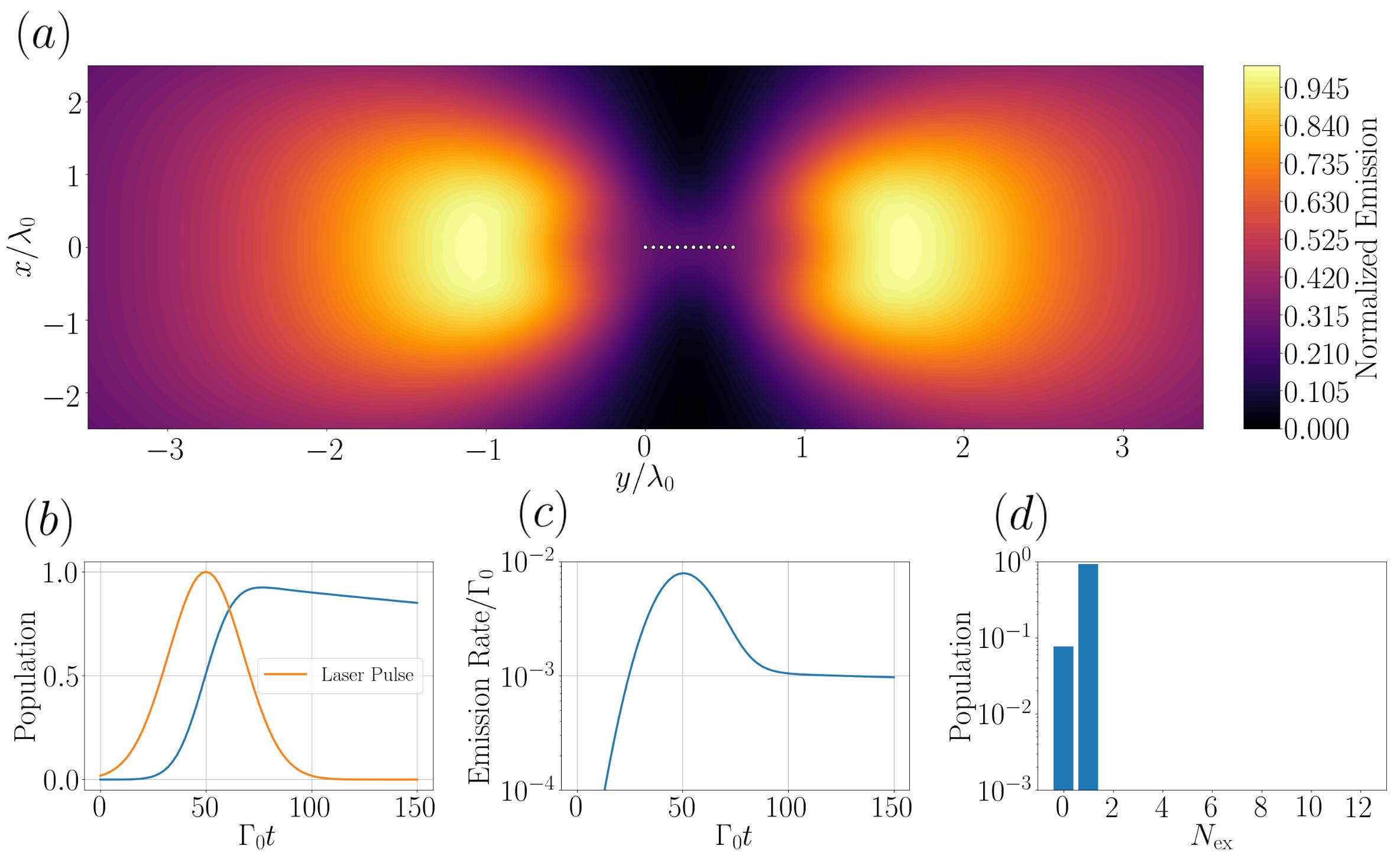}
\caption{\textit{Subradiance.} The linear chain of 12 quantum emitters along the y direction each linearly polarized in z direction with spacing $d = \lambda_0/20$. For optimal subradiant state preparation a laser pulse propagating along the chain is chosen with amplitude $\tilde{\Omega}_p = \Gamma_0$ and FWHM of the pulse duration of $\tau = 25\Gamma_0$. The laser is linear polarized and tuned to the most subradiant single excitation eigenmode of the chain. In (a) the normalized intensity distribution at $\Gamma_0 t =150$ in the xy plane is shown with a cut at $z=20d$. The emission rate in (c) shows that the chain radiates weakly only at its ends with a rate $10^{-3} \Gamma_0$. In (b) the laser pulse and the emitter population of the chain are plotted with the chain population reaching almost unity. (d) shows the distribution of the population in the individual excitation manifolds at $\Gamma_0 t =150$. Only the total ground state and the first excitation manifold are populated showing that a single excitation is stored in the chain which radiates at a rate $10^{-3}\Gamma_0$.}
\label{subradiance}
\end{figure*}

For an infinite chain that extents along the $y$ direction the eigenstates of $\mathcal{H}_{\mathrm{eff}}$ are spin waves with a well defined wave vector $k_y \in [-\pi/d,\pi/d]$. The collective spin operators $\hat{S}^\dagger_{k_y} = 1/\sqrt{N} \sum_j e^{i k_y y_j} \sigma^+_j$ satisfy the eigenvalue equation $\mathcal{H}_{\mathrm{eff}} \hat{S}^\dagger_{k_y} |g\rangle^{\otimes N} = \omega_{k_y}\hat{S}^\dagger_{k_y}|g\rangle^{\otimes N}$, where $y_j$ is the position along the chain of emitter $j$ and $|g\rangle^{\otimes N}$ the total ground state of the sytem. The complex eigenvalues read
 \begin{equation} \label{k-vector}
 \omega_{k_y} = \omega_0 - \frac{3\pi \Gamma_0}{k_0} \hat{\pmb{\mu}}^* \cdot \tilde{\pmb{G}}(k_y) \cdot \hat{\pmb{\mu}},
\end{equation}
assuming all quantum emitters have the same unit dipole moment $\hat{\pmb{\mu}}$. The imaginary part of $\omega_{k_y}$ corresponds to the decay rate of the spin wave, and its real part accounts for the frequency shift with respect to the bar atomic frequency $\omega_0$.
In the above equation $\tilde{\pmb{G}}(k_y) = \sum_j e^{-i k_y y_j} \pmb{G}(y_j,\omega_0)$ is the discrete Fourier transformation of the vacuum Green's tensor.

Fig. 1(a) shows the real and imaginary part of $\omega_{k_y}$ as well as the light line beyond which ($k_y>k_0=\omega_0/c$) the spin wave mode becomes extremely subradiant as opposed to modes inside the light line ($k_y<k_0$) which acquire a finite life time and some even become superradiant. In fact for the special case of infinitesimal emitter distances there will be $N-1$ subradiant states and a single superradiant state. It is the superradiant mode closest to the lower energy band of Fig. 1(a) which we address with a coherent driving laser such that the detuning $\Delta_l$ in the main text is on resonance with said mode. For guidance a chain of 50 emitters is shown vs. a chain of 5 emitters with a single collective mode $k_y^{(0)}$ inside the region enclosed by the light line and two subradiant modes outside of it. Due to the finiteness of the system the solution is only approximate but still a qualitative picture can be extracted as there is a clear connection between the energy band and the scattered light observed in Fig. 1(b).
As the laser is tuned to the frequency corresponding to the mode $|\Psi_{k_y^{(0)}} \rangle = 1/\sqrt{N} \sum_j e^{-i k_y^{(0)} y_j} \sigma^+_j |g\rangle^{\otimes N}$ superradiant emission ($>\Gamma_0$) is observed in the steady state total emission rate with vanishing populations in the higher excitation manifolds.
The same holds for the subradiant modes where the population is saturating at $1/2$ excitations. Note that the emission rate for $k_y^{(2)}$ amounts to $\Gamma_0/100$ in the steady state corresponding to a $100$ fold increased lifetime of the stored energy compared to independent emitters. As one increases the number of emitters in the chain these effects increase significantly with the most subradiant mode's decay rate scaling with $N^{-3}$ for large N \cite{asenjo2017exponential}. As discussed in the main text for the superradiant mode to show strong anti-bunching in the scattered light the chain length should not exceed the wavelength $\lambda_0$ of the emitted light of a single atom.

\section{Pulsed preparation of a subradiant state}

In Fig. S\ref{subradiance} we show the pulsed preparation of a subradiant single excitation eigenmode of a linear chain of 12 emitters. In Fig. 5 of the main text it is shown that a steady state operation leads to a 50$\%$ occupation of the first excitation manifold but as we show in Fig. S\ref{subradiance}(b-d) the occupation can reach near unity if the pulse duration, amplitude and detuning are optimal. The emitters are linear polarized in z direction and the chain extents along y with lattice constant $d = \lambda_0/20$ and with a linear polarized laser pulse propagating along the chain.
The Hamiltonian in Eq. 1 of the main text will now have a time dependence as follows

\begin{dmath}
   \mathcal{H}(t) = \sum_j^N \Delta_p \sigma^+_j \sigma^-_j + \sum_{i,j:i \neq j}^N \Omega_{ij} \sigma^+_i \sigma^-_j + \\ +\Omega_p(t) \pmb{\epsilon}_p \cdot \sum_j^N \hat{\pmb{\mu}}_j \left(e^{-i \pmb{k} \cdot \pmb{r}_j} \sigma^+_j + e^{i \pmb{k} \cdot \pmb{r}_j} \sigma^-_j \right),
   \label{eq:hamiltonian2}
\end{dmath}

where $\Omega_p(t)= \tilde{\Omega}_p e^{-(t-t_0)^2/\tau^2}$ is a laser pulse with amplitude $\tilde{\Omega}_p$, a full width half maximum of the pulse duration of $\tau$ and peak at $t_0$. In Fig. S\ref{subradiance}(b) the case of a laser pulse with $\tau = 25\Gamma_0$, $\Omega_p = \Gamma_0$ and $t_0 = 50\Gamma_0$ is plotted. The laser is now tuned to the most subradiant single excitation eigenmode of the chain. In (a) the normalized intensity distribution is plotted at $\Gamma_0 t = 150$ and shows that the chain radiates only at the ends with a emission rate of $10^{-3}\Gamma_0$ as is shown in (c). The distribution of the emitter population in the individual excitation manifolds in (d) at $\Gamma_0 t = 150$ shows that the chain stores nearly a single excitation which decreases at a rate of $10^{-3}\Gamma_0$. Experimentally verifying the preparation of subradiant states has always been a challenge and this should only illustrate the preparation with a minimal amount of elements.

\section{Nano Rings}

 As another example of ordered arrays the ring provides closed boundary conditions, namely rotation symmetry thereby allowing for an explicit calculation of eigenmodes.\cite{moreno2019subradiance} As in the case of the chain the Hamiltonian can be cast into a non Hermitian form where the complex eigenvalues read

\begin{figure*}[ht]
\includegraphics[width=0.95\textwidth]{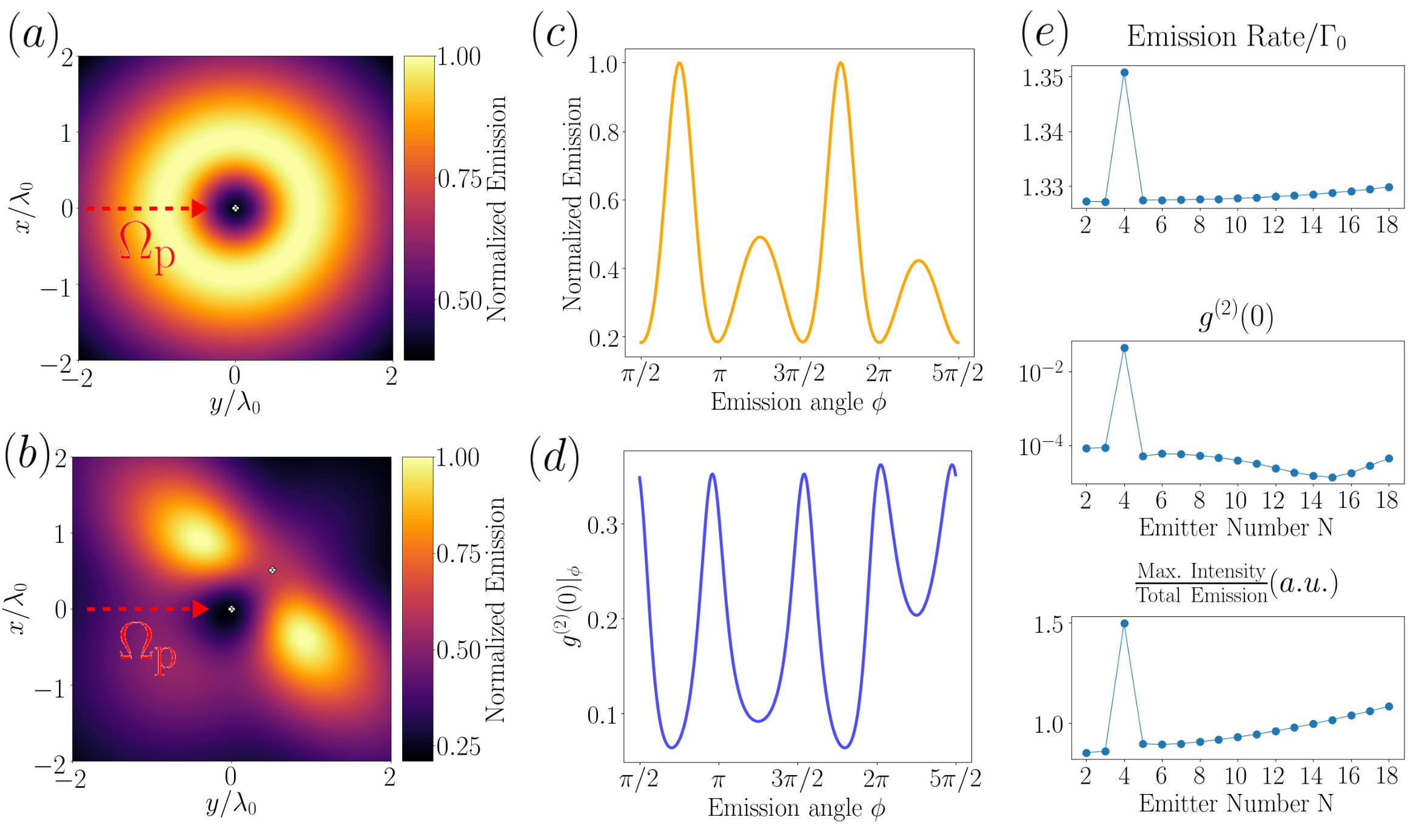}
\caption{\textit{Nano Ring.} (a) A single ring of emitters ($N=4$) is continuously driven by a linear polarized laser from the y direction with driving rate $\Omega_p = \Gamma_0$. The emitters are linear polarized in z direction with a separation $d = 0.02 \lambda_0$. The laser is tuned to the most superradiant single excitation eigenmode of the ring which is given by $\omega_0 + \sum_{j=2}^N \Omega_{1j}$ but steady state emission shows no directionality in the xy-plane. (Cut in $z = 1.5\lambda_0$) In (b) a second identical undriven ring is placed at an $\pi/4$ angle to the first ring with the centers of the rings being $0.7 \lambda_0$ apart and the steady state showing strong directional emission in a direction which is not interfering with the laser beam direction. The intensity maxima in (b) are shown in (c) as a function of the emission angle $\phi$ (xy-plane) and (d) shows correspondingly a low bunching parameter $g^{(2)}(0)<0.1$ in the directions of maximum emission. Scaling behaviours as a function of the undriven ring emitter number are plotted in (e) for the total emission rate, $g^{(2)}(0)$ in the direction of maximum emission and the ratio between maximal emission and total emission in the steady state. The emitter number of the driven ring is fixed at $N=4$ as is the distance between the ring centers. For an increasing emitter number in the undriven ring, more light is concentrated in the direction of maximal emission with a $g^{(2)}(0) \le 10^{-3}$ as seen in (e), constituting a nearly perfect directional single photon source.}
\end{figure*}

 \begin{equation}
 \label{ring_omega}
 \omega_{m} = \omega_0 - \frac{3\pi \Gamma_0}{k_0} \hat{\pmb{\mu}}^* \cdot \tilde{\pmb{G}}(m) \cdot \hat{\pmb{\mu}},
\end{equation}

where $\tilde{\pmb{G}}(m) = \sum_{jl} e^{-i m (\varphi_l-\varphi_j)} \pmb{G}(\pmb{r}_l-\pmb{r}_j,\omega_0)$ is again the discrete Fourier transformation of the Green's tensor for the ring. The angle associated with position $j \in (1,...,N)$ is denoted by $\varphi_j = 2\pi (j-1)/N$ and $m = 0,\pm 1,\pm 2,..., \lceil \pm(N-1)/2\rceil$ corresponds to the angular momentum of the mode.\cite{moreno2019subradiance,Cremer_2020}
The collective energy shifts and emission rates of the mode are given by the real part $\Omega_m = \mathrm{Re}\{\omega_m\}$ and the imaginary part $\Gamma_m = \mathrm{Im} \{\omega_m\}$ of the eigenvalue $\omega_m$.
From Eq. \ref{ring_omega} it is easy to see that the eigenvalue spectrum will be symmetric under the exchange $m \leftrightarrow -m$, meaning, $\omega_m = \omega_{-m}$. Relevant for the present case is the superradiant $m = 0$ mode which is non-degenerate and will be targeted by a coherent pump of rate $\Omega_p = \Gamma_0$.
The corresponding eigenstate has the form
 \begin{equation} \label{eq:ring1}
|\Psi_{m=0} \rangle = \frac{1}{\sqrt{N}}\sum_{j=1}^N \sigma^+_j |g\rangle^{\otimes N}
\end{equation}
with a superradiant emission rate $\Gamma_{\mathrm{sup}} = \sum_{j=1}^N \Gamma_{1j}$ and eigenenergy $\omega_0 + \sum_{j=2}^N \Omega_{1j}$ which has the same form as the eigenenergies of the hermitian Hamiltonian which is used for the simulations.
Note that for infinitesimal emitter spacing the ensemble approaches the Dicke Limit\cite{dicke1954coherence} for which the superradiant mode decays with rate $N \Gamma_0$.
Now we will investigate a coherently driven ring of emitters of separation $d=0.02\lambda_0$ each having linear polarization in z direction with a linear polarized laser propagating in the y direction with a pumping rate $\Omega_p = \Gamma_0$ (see Eq. 1 in the main text). Fig. S3(a) shows the normalized intensity distribution in the steady state for $N =4$ emitters in the ring and features a uniform ring shaped emission into the xy-plane. Now in (b) a second identical undriven ring is placed at a $\pi/4$ angle w.r.t. the first ring with a separation between the two rings centers of $0.7 \lambda_0$. The resulting steady state emission shows strong directionality in the xy-plane in a direction which is not interfering with the laser's direction. The 1D plot in (c) shows the corresponding emission peaks in the far field and (d) the bunching parameter $g^{(2)}(0)$ as a function of the emission angle $\phi$ in the xy-plane. A bunching parameter of $<0.1$ can be observed in the direction of maximal emission coinciding exactly with the global minima of the $g^{(2)}(0)$ function which would constitute a good single photon source. Finally in (e) the scaling behaviour is plotted when the emitter number in the undriven ring is varied leaving all other parameters fixed including the distance between the two ring centers.
The steady state total emission rate remains roughly constant with a small variation at $N=4$. This is expected as the driven ring's emitter number remains constant and which is the source of the emission or scattered light. The bunching parameter on the other hand shows strong variations of two orders of magnitude as the second ring's emitter number is changed. The ratio between the maximal emission in the $xy$-plane and the the total emission in (d) is increasing with increasing emitter, therefore more light is concentrated in the direction of maximal emission. Simultaneously a bunching parameter $g^{(2)}(0) \approx 10^{-4}$ is reached which would constitute an almost perfect source of single photons. Although driving only one of the rings might be a challenge experimentally the small bunching parameter in combination with the strong directionality as seen in Fig. S3(b,c) and a total emission rate of $>\Gamma_0$ in this system seems to be an interesting avenue for future experimental and theoretical studies.

\section{Tilted Polarization}

To drive only one of the rings in the previous example might prove experimentally difficult. By again considering the same geometry as in Fig. S3 but adding a small x component of $\epsilon = 0.1$ to the polarization of the left ring and the laser being circular polarized as $(\hat{x},i \hat{y} ,0)^T/\sqrt{2}$ and propagating along the z direction.
With a pumping rate of $\Omega_p = 20\Gamma_0$ the laser beam will only drive the left ring which still has a dipole-dipole interaction with the second ring. The normalized steady state emission for $N=14$ emitters per ring is plotted in Fig. S4 (b) and shows strong directionality in the $xy$-plane. In (a) the steady state total emission rate as a function of the emitter number per ring is shown in a, whereas in (c) and (d) the bunching parameter $g^{(2)}(0)$ in the direction of maximal emission and the ratio between maximal emission and total emission in the $xy$-plane are plotted. At $N=14$ the whole system shows a total emission rate approximately $7$ times larger than that of a single emitter which is $\Gamma_0/2$, a bunching parameter $g^{(2)}(0) \approx 0.03$ and increasingly more light concentrated in the direction of maximal emission. In effect the tilted polarization of the driven ring converts the circular polarized laser light into highly antibunched linear polarized light in a continuous manner.

 \begin{figure}[ht]
\includegraphics[width=0.49\textwidth]{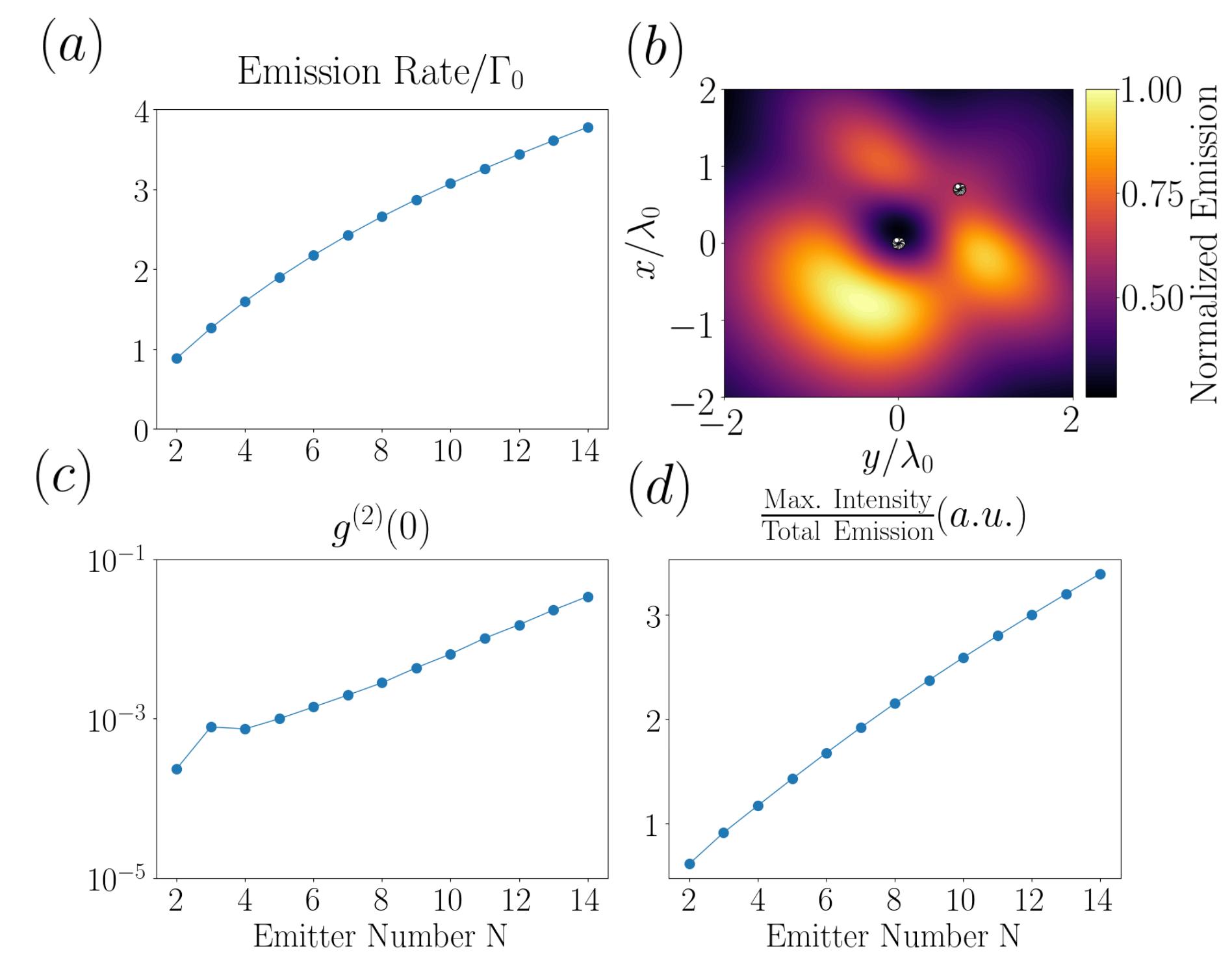}
\caption{\textit{Tilted Polarization}. Taking the same arrangement as in Fig. S3 but adding a small component of $x=0.1$ to the dipole orientations ($\hat{\pmb{\mu}}=(0,0,1)^T$) of the emitters in the driven ring on the left. The laser beam propagates in the z-direction with a rate $\Omega_p = 20\Gamma_0$ and a circular polarization $(\hat{x},i \hat{y} ,0)^T/\sqrt{2}$.
In (b) the normalized steady state emission rate in the xy-plane is shown for $N=14$ emitters per ring (Cut in $z = 1.5\lambda_0$). The total emission rate, the bunching parameter and the ratio between maximal emission and total emission in the steady as a function of the emitter number per ring are shown in (a), (c) and (d) respectively.
A single ring of such small scale with transversally polarized emitters would emit into the whole xy plane equally but the presence of a second identical undriven ring directs the emission into a particular direction. }
\label{tilted}
\end{figure}

\section{Spatial Disorder}

\begin{figure}
\includegraphics[width=0.99\columnwidth]{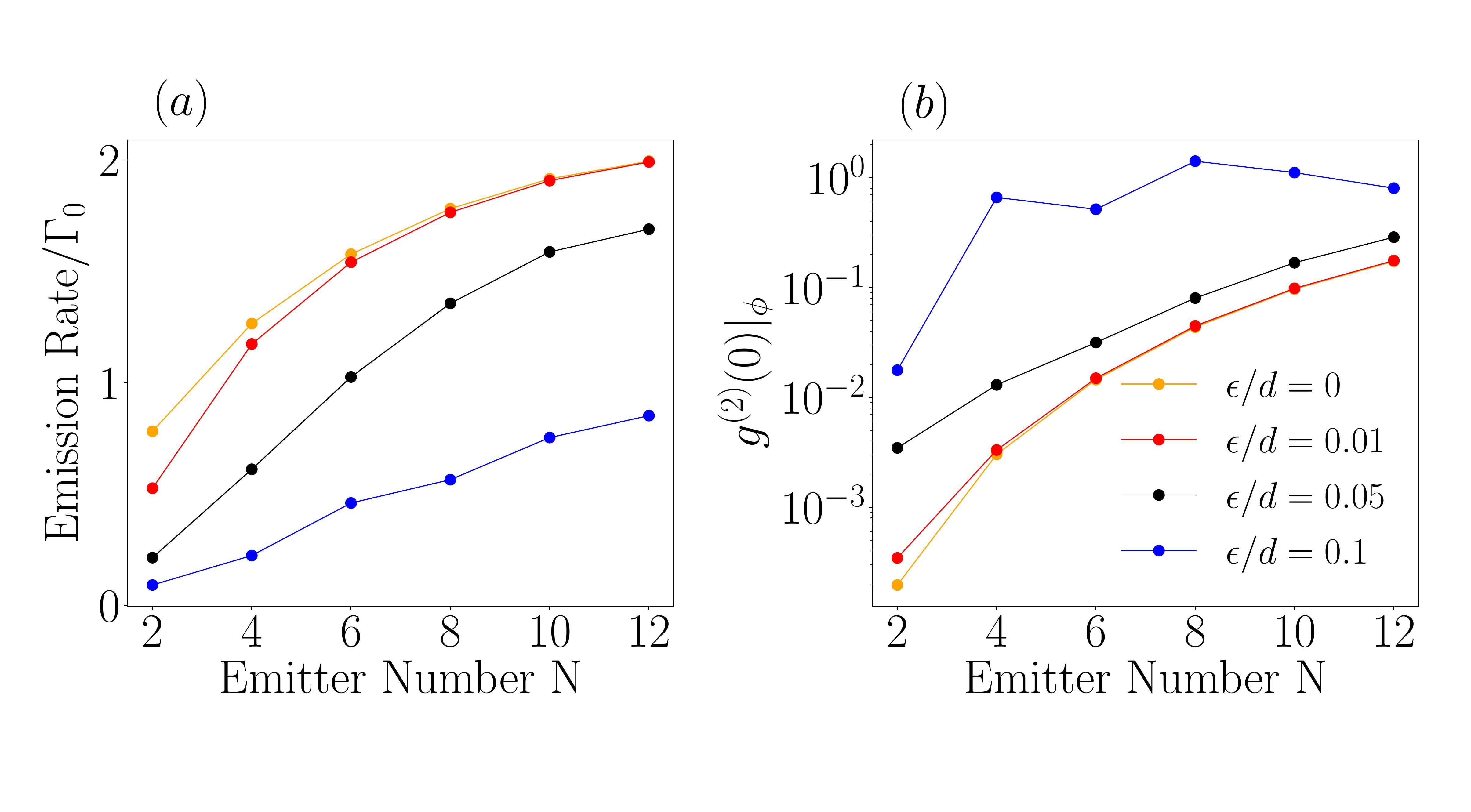}
\caption{Influence of positional disorder $\epsilon$ for a linear chain in (a) the steady state emission rate and (b) bunching parameter $g^{(2)}(0)$ in the direction of maximal emission.
The continuous lines are guides for the eye showing the dependence on the emitter number N. For a chain along $y$, each emitter is displaced in $x$ and $y$ by a random value between $ \ [-d\epsilon,d\epsilon]$. We average over 100 random configurations for each value of $\epsilon$. For all plots, $d=\lambda_0/40$.
}
 \label{disorder}
\end{figure}

In Fig. S\ref{disorder} we plot in (a) the total emission rate and in (b) the bunching parameter $g^{(2)}(0)$ in the direction of maximal emission as a function of the emitter number in a linear chain along y for different degrees of spatial disorder. Each emitter is linear polarized in z direction and the linear polarized driving laser propagates along the chain direction with rate $\Omega_p = \Gamma_0$. The laser is tuned to the most superradiant single excitation state for each N where $\epsilon=0$. This means that the positional disorder also induces some detuning between the laser frequency $\omega_p$ and the targeted eigenmode of the disordered chain. We consider a classical disorder, where each emitter is randomly displaced around its initial position in the $xy$ plane by a value between $ \ [-d\epsilon,d\epsilon]$.
We plot the total emission rate and the bunching parameter after averaging over 100 disorder realizations. The chain lies along y and the disorder for each emitter is in x and y direction. Evidently in the presence of significant fluctuations of $\epsilon/d = 0.1$ the total emission rate is substantially decreased and the bunching parameter increased on the other hand moderate disorder has only small effects on both values. In the case of subradiance it was shown that moderate disorder has relatively little influence on the decay rate of the most subradiant eigenmode. \cite{asenjo2017exponential}

\section{Truncated Model}

Troughout this work a truncated Hilbert space is used for most simulations by which the full quantum model is restricted only up to two excitations.
In this way it is still possible to calculate second order correlations in normal order of the form $\langle \hat{\sigma}^+_i \hat{\sigma}^+_j \hat{\sigma}^-_k \hat{\sigma}^-_l \rangle$ and investigate larger system sizes with finite computational resourcres.

In order to calculate the occupation in a given excitation manifold we have diagonalized the given Hamiltonian and summed up the projections of the steady state density matrix on the respective eigenstates. For instance in the first excitation manifold are $N$ eigenstates, so the population is given by $\sum_{j=1}^N \langle \psi_j | \rho_{ss}|\psi_j\rangle$, where $|\psi_j\rangle$ denote the $N$ single excitation eigenstates and $\rho_{\mathrm{ss}}$ is the density matrix for the emitters in the steady state. There are $N(N-1)/2$ eigenstates carrying two excitations and so forth with all excitation manifolds adding up to $2^N$ states.

 \begin{figure}[ht]
\includegraphics[width=0.49\textwidth]{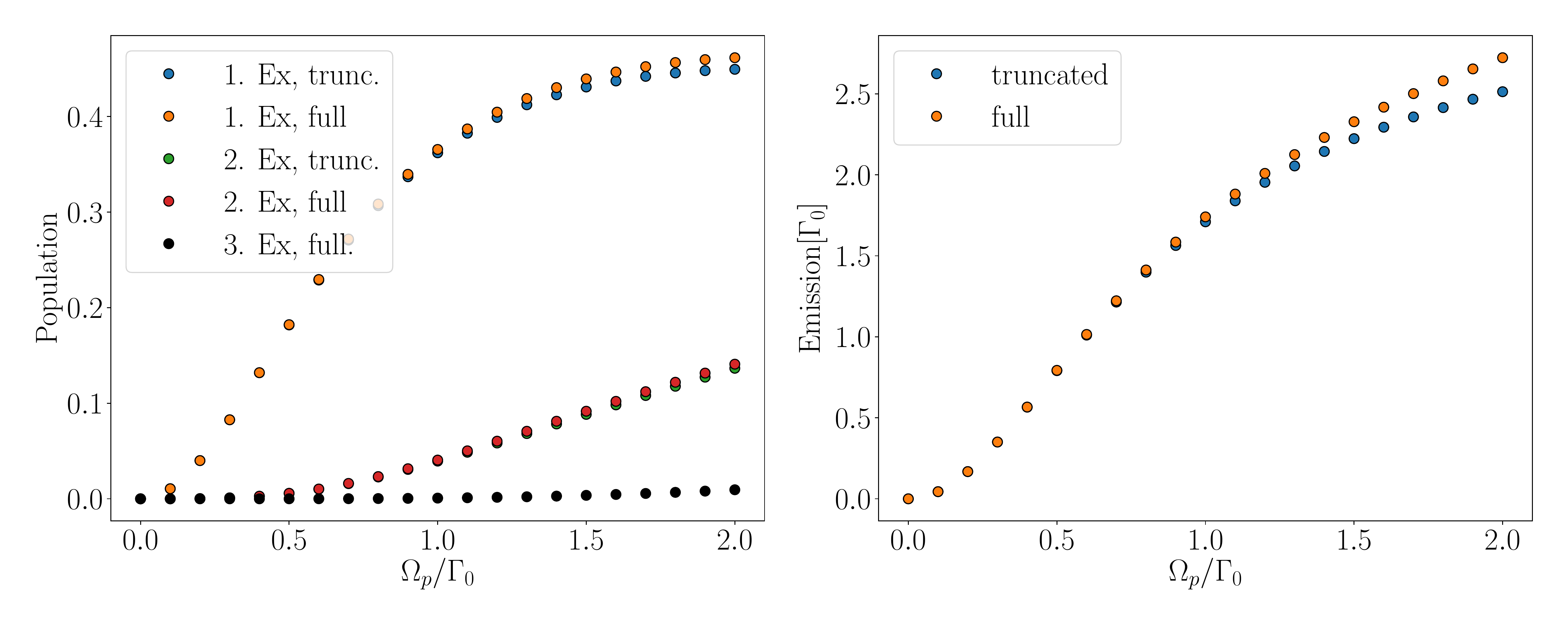}
\caption{A comparison of the full quantum model with the truncated model is shown in the steady state for a 5 emitter chain along y with spacing $d/\lambda_0 = 0.05$ and the emitters linearly polarized in z direction. The laser is again tuned to the most superradiant single excitation eigenmode and propagates along the chain direction. The truncation includes only up to two excitation but it is seen that the third excitation manifold in the full model is almost negligible and the total emission rate shows good agreement up to a coherent driving rate of $\Gamma_0$ both for the steady state population and the total emission rate.}
\label{full-truncated}
\end{figure}

In Fig. S5 the full quantum model is compared to the truncated model for a linear chain of 5 emitters both for in population and total emission rate versus pumping rate $\Omega_p$.
A good approximation to the full model for small enough coherent pumping rates $\Omega_p \le \Gamma_0$ can be found. The vanishing population in the third excitation manifold of the full model shows that the truncated model describes the system sufficiently well from a physics viewpoint.

\end{document}